\documentclass{appolb}
\usepackage{graphicx,subfigure,amsmath,amssymb}
\begin{document}
% \eqsec  % uncomment this line to get equations numbered by (sec.num)
\title{The $T-\mu$ phase diagram of the NJL model in the presence of explicit symmetry-breaking interactions
%\thanks{Presented at \textit{EEF70 - Workshop on Unquenched Hadron Spectroscopy: Non-Perturbative Models and Methods of QCD vs. Experiment.}}
% you can use '\\' to break lines
}
\author{J. Morais, J. Moreira, B. Hiller, A. H. Blin, A. A. Osipov
\address{Departamento de F\'isica da Universidade de Coimbra, 3004-516, Coimbra, Portugal}
}
\maketitle
\begin{abstract}
It is shown that the strange quark mass undergoes a first order transition in a generalized 3 flavor Nambu--Jona-Lasinio (NJL) Lagrangian which  includes a complete set of explicit chiral symmetry breaking interactions. This transition occurs in a moderate chemical potential region $\mu \sim 400$ MeV, in addition to the usual chiral transition associated with the light quark sector. This favors the formation of stable strange quark matter at chemical potentials which are considerably lower than the ones discussed in the literature. The reason for this behavior is discussed.
\end{abstract}
  
\section{Introduction - QCD Phase Diagram}
Currently, the QCD phase diagram is a subject under intense investigation, both theoretically and experimentally. To this date, most of what has been said about its structure is highly conjectural. There are but a few features of the QCD phase diagram which may be regarded as being consensual among different approaches to its description, for a revew see for instance \cite{Fukushima:2010bq}. A particular aspect of strongly interacting matter that remains as a hypothesis is the existence of strange quark matter (SQM).% The development of a more accurate description of the QCD phase diagram will certainly clarify this issue.

At asymptotic high densities, QCD can be studied perturbatively. Such a treatment predicts color superconducting phases at low to moderate temperatures, characterized by a finite diquark condensate. However, at lower energies and density, QCD must be studied non-perturbatively,
%. Therefore, the lower density region of the QCD phase diagram, which possibly exhibits a complex structure, is mostly investigated 
for instance through lattice calculations and effective models. There is a general agreement that a first-order boundary exists between hadronic and quark matter phases, as well as a critical endpoint for this boundary~\cite{Fukushima:2010bq}. The physical mechanisms which appear to be relevant for the existence of such a boundary are color confinement and chiral symmetry breaking, with the phase structure emerging from their interplay. An approximate match between deconfinement and chiral transitions is commonly assumed. The existence of a critical endpoint is supported by the lattice result that the hadron-quark matter boundary assumes the form of a smooth crossover for $\mu=0$~\cite{Fukushima:2010bq,Philipsen:2011zx}.

The hot region of the QCD phase diagram may be probed in heavy ion collisions, as has been done at Brookhaven's RHIC. Besides producing data relevant to the phase diagram, experiments at RHIC have searched for the creation of strangelets, although no signs of their existence have been found~\cite{Adams:2005cu}. The cold dense region is of relevance to compact stars, where it is also speculated that SQM might form~\cite{Weber:2004kj}. Observational data from these stellar objects may be used as an empirical constraint to the description of strongly interacting matter.

\section{Thermodynamic Potential and Phase Diagram}
The starting point of the present study is a three-flavor NJL-type effective Lagrangian with the complete set of spin-0 non-derivative multi-quark interactions which contribute up to next to leading order in $N_c$ counting: the current mass  and 4q terms of the NJL model (both of which are leading order), the 't Hooft 6q determinantal term, two 8q terms and additional current quark mass dependent interactions which explicitly break chiral symmetry~\cite{Osipov:2013fka}. The model is bosonized in the functional integral formalism, using a stationary phase approximation for the integration over auxiliary bosonic degrees of freedom, and a generalized heat kernel expansion for the integration over quark degrees of freedom. The effective potential is then determined through standard techniques in the mean field approximation:
\begin{align}
\mathcal{V} & = \left[ \frac{G}{4} \sum_i h_i^2 + \frac{\kappa}{2} h_u h_d h_s + \frac{3 g_1}{16} \left( \sum_i h_i^2 \right)^2 + \frac{3 g_2}{16} \sum_i h_i^4 \right. \nonumber \\
 & + \frac{\kappa_2}{2} \sum_{i \neq j \neq k} m_i h_j h_k + \frac{g_3}{4} \sum_i m_i h_i^3 + \frac{g_4}{4} \sum_i h_i^2 \sum_j m_j h_j \label{eq-eff-pot}\\
 & \left.\left. + \frac{g_5 + g_6}{8} \sum_i m_i^2 h_i^2 + \frac{g_7}{4} \left( \sum_i m_i h_i \right)^2 \right]\right|_0^{M_i} + \frac{N_c}{8\pi^2} \sum_{i=u,d,s} J_{-1} \left(M_i^2\right) \nonumber
\end{align}
\noindent In~(\ref{eq-eff-pot}), $G$, $\kappa_n$ and $g_n$ are model parameters, $\Lambda$ is a regularization scale, $m_i$ are quark current masses, and $h_i$ are twice the quark-antiquark condensates. The $J_{-1} \left(M_i^2\right)$ are integrals which stem from the quark heat kernel, $M_i$ being the quark dynamical masses. Temperature $T$ and chemical potentials $\mu_i$ are included through these integrals in the Matsubara formalism, giving a thermodynamic potential of the form:
\begin{equation} \label{eq-therm-pot}
\Omega \left(T,\mu_i\right) = \mathcal{V}_{st} + \frac{N_c}{8\pi^2} \sum_{i} J_{-1} \left(M_i^2,T,\mu_i\right) + C \left(T,\mu_i\right),
\end{equation}
\noindent where $\mathcal{V}_{st}$ is the term in brackets in \ref{eq-eff-pot}, and $C$ is a term which is fixed by the correct asymptotic behavior of $\Omega$.

Quark dynamical masses $M_i$ are obtained by numerically solving the gap equations subject to the stationary phase conditions:
\begin{align}
\label{eq-main}
h_i &= - \frac{N_c}{2\pi^2} M_i J_0 \left( M_i^2, T, \mu \right)\nonumber\\
M_i & = m_i - \frac{h_i}{2} \left( 2G + g_1 h^2 + g_4 mh \right) - \frac{g_2}{2} h_i^3  - \frac{\kappa}{4} t_{ijk} h_j h_k - \frac{\kappa_2}{2} t_{ijk} m_j h_k \nonumber \\
& - \frac{m_i}{4} \left( 3 g_3 h_i^2 + g_4 h^2 + \left(g_5 + g_6 \right) m_i h_i + 4 g_7 mh \right)
\end{align}

\begin{figure*}
\center
\subfigure[]{\label{grafMuMsmu0300}\includegraphics[width=0.32\textwidth]{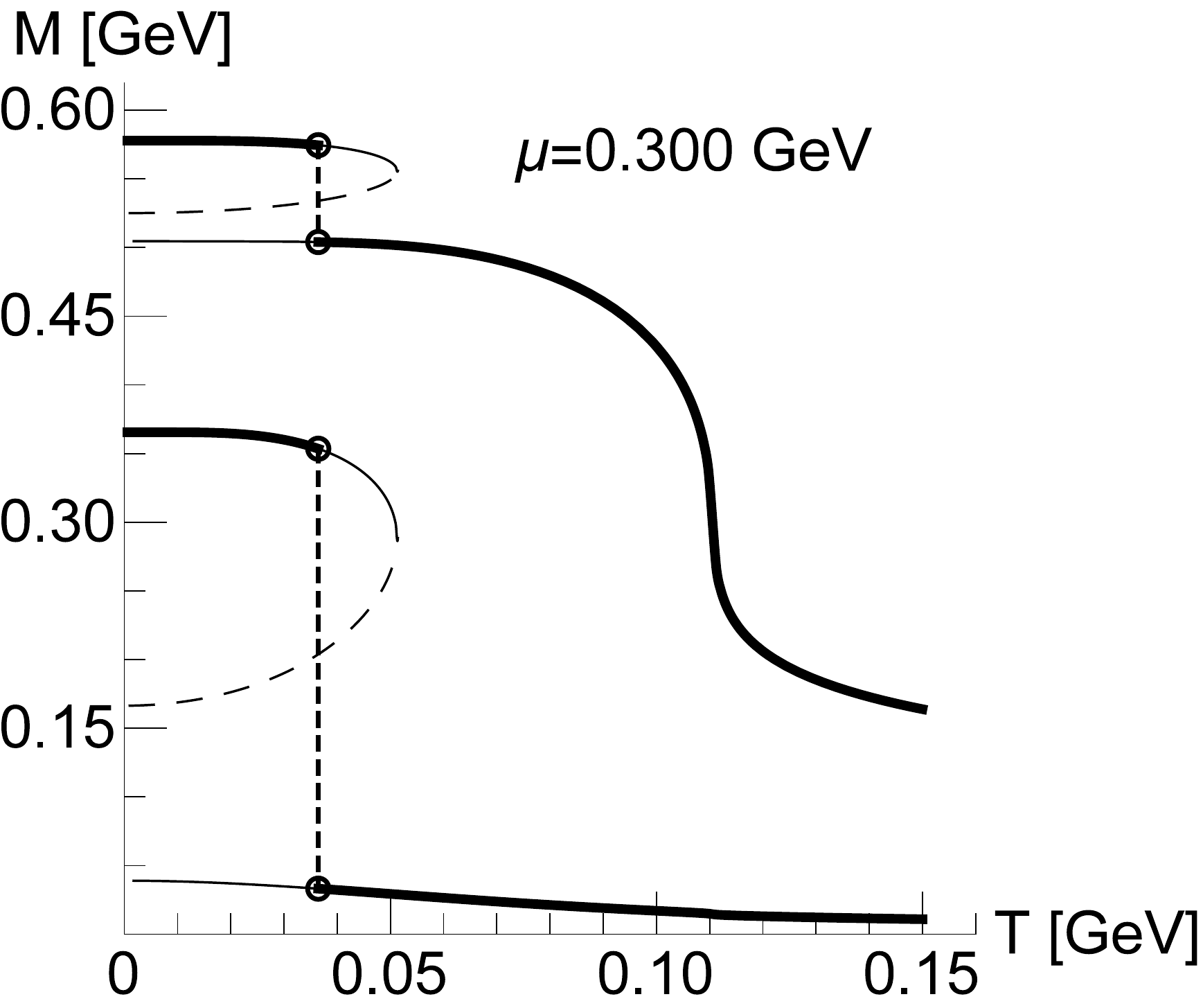}}
\subfigure[]{\label{grafMuMsmu0400}\includegraphics[width=0.32\textwidth]{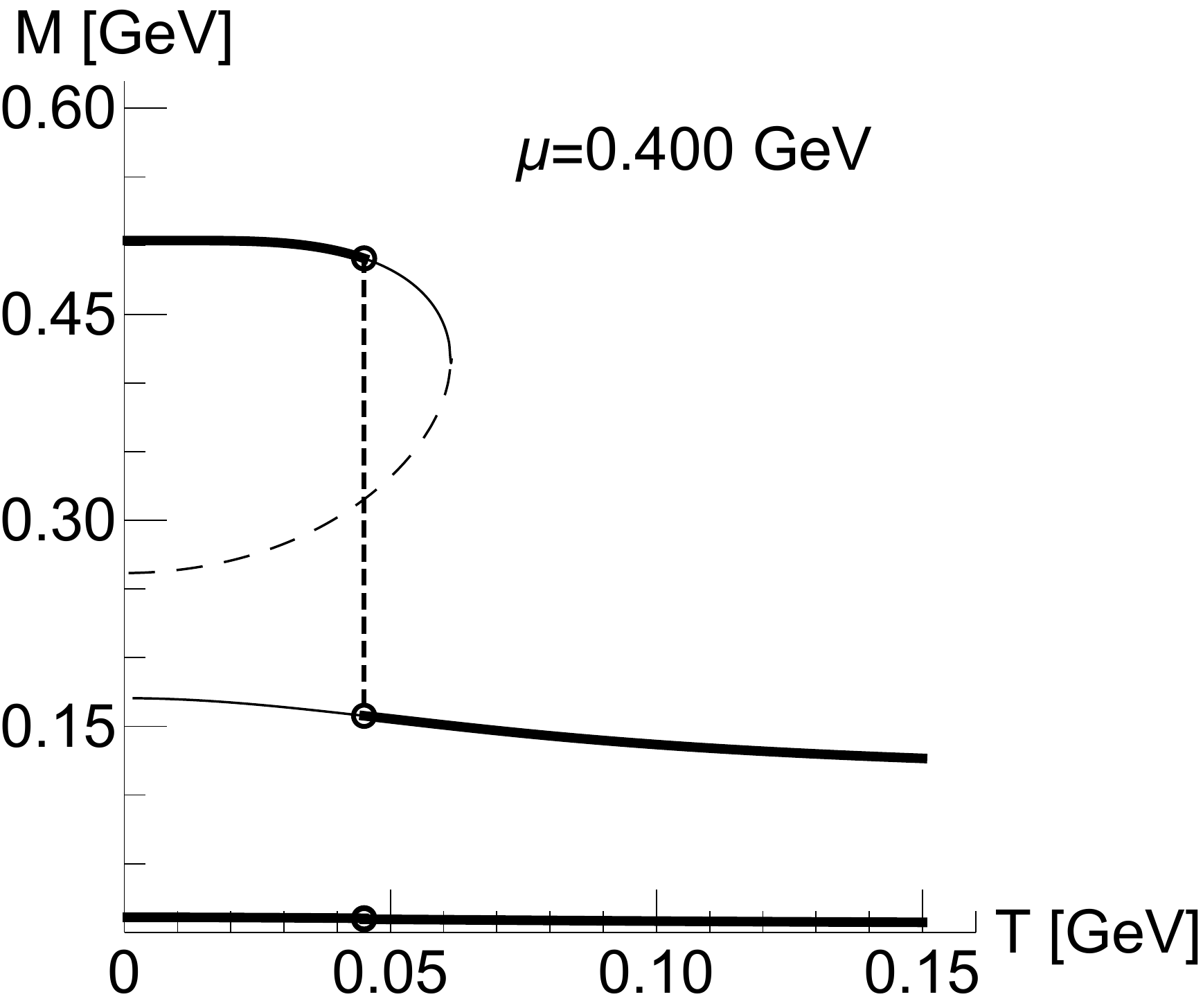}}
\subfigure[]{\label{grafPDD2}\includegraphics[width=0.32\textwidth]{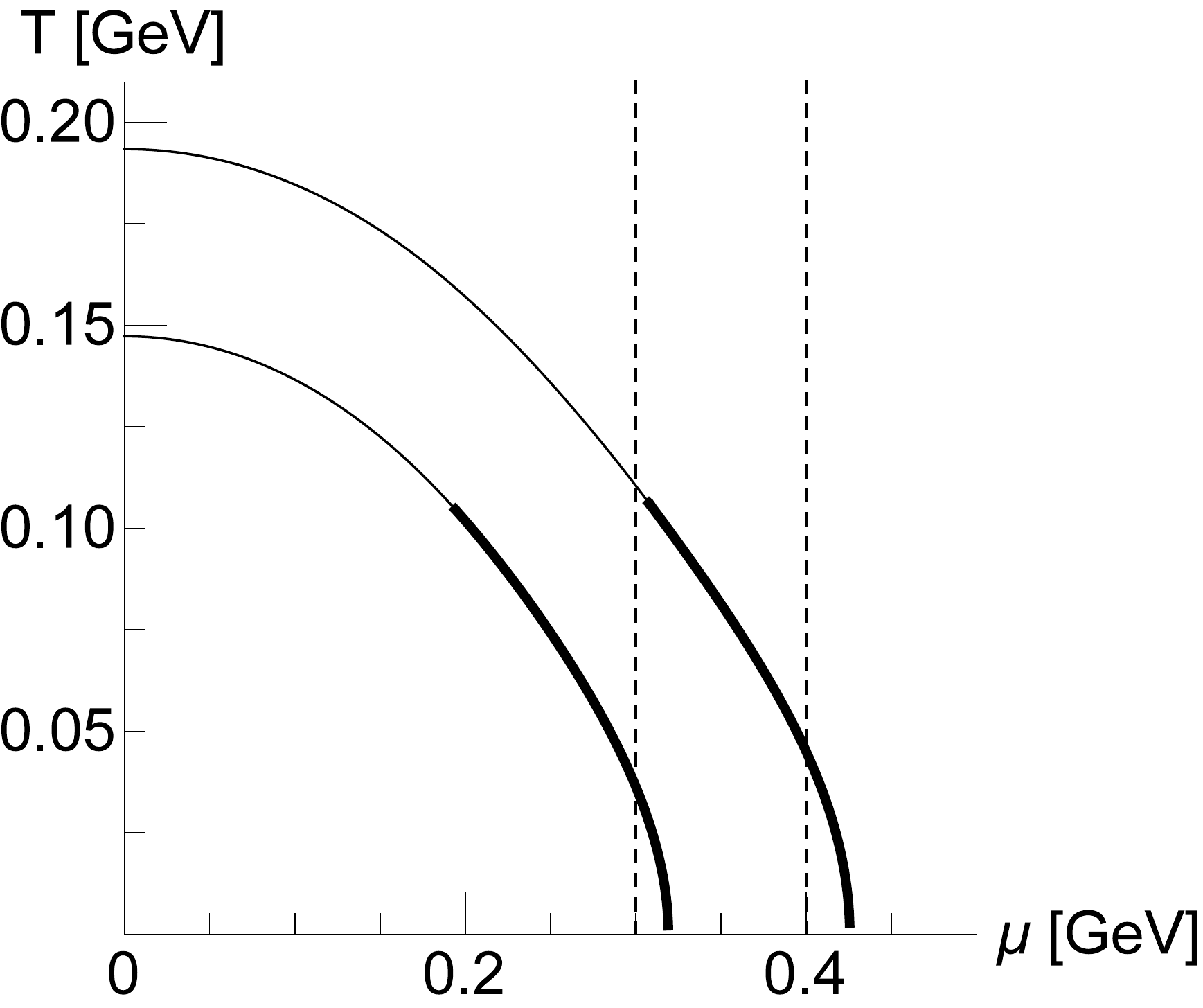}}
\caption{
Dynamical mass profiles as a functions of temperature at the two chemical potentials indicated in \ref{grafPDD2} by the vertical lines: $\mu=0.300~\mathrm{GeV}$ in Fig. \ref{grafMuMsmu0300} and $\mu=0.400~\mathrm{GeV}$ in Fig. \ref{grafMuMsmu0400}. The upper two lines are for $M_s$, the lower ones for $M_l$; thick lines represent physical solutions, thin and dashed lines show relative minima and maxima of the thermodynamic potential, respectively;  first-order transitions are represented as dashed lines connecting two circles. Phase diagram in the $\mu-T$ plane using the parameter set from \cite{Moreira:2014qna} displaying the two transition lines (first order/crossover corresponds to the thick/thin lines) associated with the light (inner lines) and strange quarks (outer lines) in \ref{grafPDD2}.  In these figures, $\mu_u = \mu_d = \mu_s = \mu$.
}
\label{Imagens_PhaseDiagram}
\end{figure*}

\noindent These solutions correspond to extrema of the thermodynamic potential~(\ref{eq-therm-pot}); minimum solutions are stable (or metastable if the minimum is local) while maximum solutions are unstable. We perform our calculations in the isospin limit ($m_u = m_d \neq m_s$) and with an average quark chemical potential $\mu_i = \mu$. A sequential analysis of dynamical mass profiles $M_i \left(T\right)$ at fixed values of $\mu$ shows a smooth crossover behavior which gets progressively steeper and shifted towards lower values of $T$ as higher values of $\mu$ are considered. At some critical value of $\mu$ we observe the onset of a first-order transition which is more pronounced in the light quark masses than in the strange quark mass (see fig. \ref{grafMuMsmu0300}). As we further increase $\mu$, a second first-order transition eventually emerges which is, in this case, substantial in the strange quark mass but almost negligible in the light quark masses (see fig. \ref{grafMuMsmu0400}). This additional first-order transition is a suprising new feature which is absent in NJL models without the new mass-dependent terms except for unphysical values of the parameters or for the case where diquark interactions are included. In fig.~\ref{grafPDD2}, we present the phase diagram in the $T-\mu$ plane, where the two first-order boundaries are shown as well as the pseudocritical points associated with the crossover regions (these are determined from the points of maximal slope of the dynamical mass functions).

\section{$\beta$-Equilibrium and Charge Neutrality at $T=0$}

In order to further understand the implications of the model for SQM, we impose $\beta$-equilibrium conditions as well as charge neutrality, and we study the thermodynamical features of the model at $T=0$. Upon neglecting neutrino contributions, these conditions may be written as:
\begin{equation}
\label{eq-beta}
\mu_u = \mu - \frac{2}{3} \mu_e, \qquad \mu_d = \mu_s = \mu + \frac{1}{3} \mu_e,\qquad 
\frac{2}{3} \rho_u - \frac{1}{3} \left( \rho_d + \rho_s \right) - \rho_e = 0,
\end{equation}
\noindent where $\mu$ and $\mu_e$ are the average quark chemical potential and the electron chemical potential, respectively, and $\rho_i$ are number densities of the species. Such a low temperature regime is relevant for compact stellar objects, which are conjectured to host favourable conditions for the formation of SQM in their cores ~\cite{Weber:2004kj}. Conditions~(\ref{eq-beta}) are appended to equations~(\ref{eq-main}) and the system is solved numerically for the dynamical quark masses $M_i$. The two critical points occur at $\mu \sim 325 \text{MeV}$ and $\mu \sim 409 \text{MeV}$.

\begin{figure*}
\center
\subfigure[]{\label{grafDensT0000EqBeta}\includegraphics[width=0.32\textwidth]{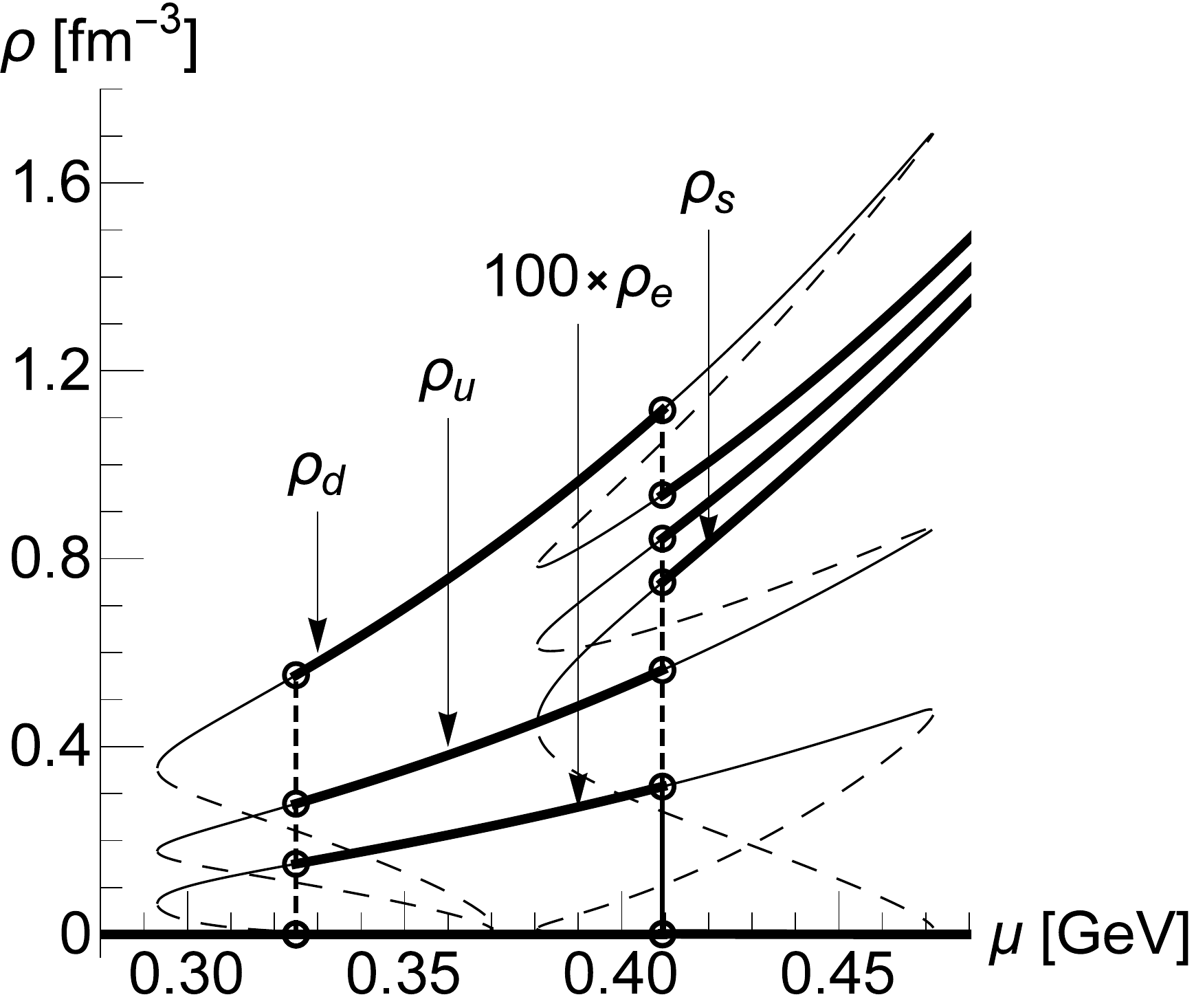}}
\subfigure[]{\label{grafEpart}\includegraphics[width=0.32\textwidth]{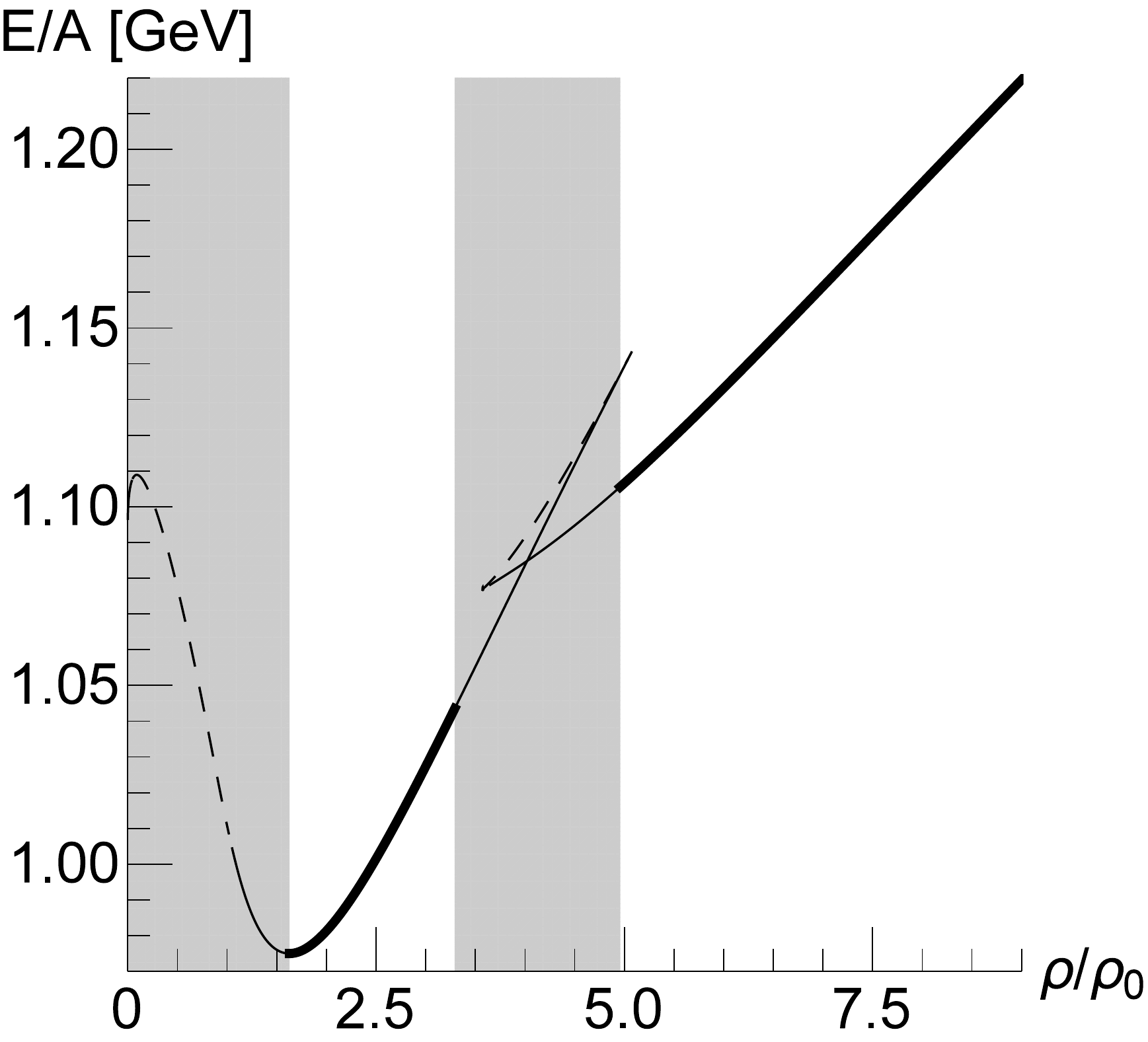}}
\subfigure[]{\label{grafpEner}\includegraphics[width=0.32\textwidth]{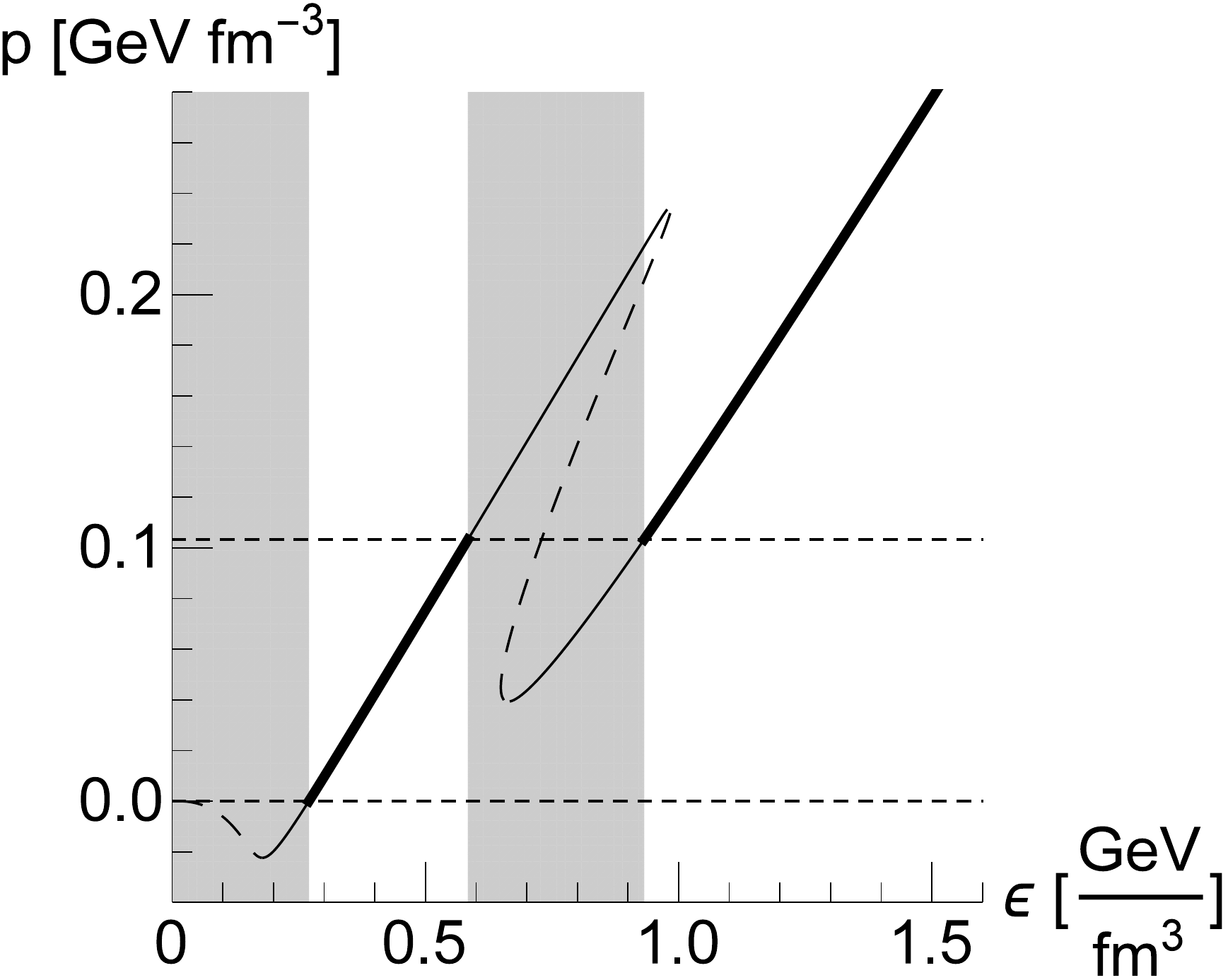}}
\caption{
Number densities for quarks ($u$, $d$ and $s$) and electrons (scaled by a factor of 100) as a functions of the chemical potential $\mu$ are shown in Fig. \ref{grafDensT0000EqBeta}.  In Fig. \ref{grafEpart} the energy per baryon number, $E/A=\epsilon/\rho$, as a function of the baryon number density $\rho=(\rho_u+\rho_d+\rho_s)/3$ (normalized to the nuclear matter saturation density $\rho_0 = 0.17 \text{fm}^{-3}$). The state equation (pressure as a function of energy density) is shown in Fig. \ref{grafpEner}. 
The gray areas correspond to a range where the system is in a mixed phase. In the lower density window the mix involves the phases with $\rho/\rho_0=0 - 1.636$ ($p=0~\mathrm{GeV~fm^{-3}}$) and in the other window it involves the phases with $\rho/\rho_0= 3.309- 4.982$ ($p=0.104~\mathrm{GeV~fm^{-3}}$).
The line notation (thin, thick and dashed) is the same as in previous figures.
}
\label{Imagens_EqBeta}
\end{figure*}

We look at the number densities of all the species as functions of $\mu$, whose stable solutions can be seen in fig. \ref{grafDensT0000EqBeta}. All densities are null up to the first transition, at which point finite up and down quark densities emerge, as well as a small electron density. These increase with increasing $\mu$ until the second transition, when a finite strange quark density appears. After this point, the electron density drops rapidly to zero, and the quark densities rise monotonically and get progressively close to each other. This latter feature is commonly pointed out as a necessary condition for the formation of SQM~\cite{Buballa:2003qv}.

Using standard thermodynamic relations (at $T=0$) for the energy density $\epsilon$ and the number densities $\rho_i$, we also study how the energy per baryon $E/A = \epsilon / \rho$ varies with the relative baryonic density $\rho / \rho_0$, with $\rho = (\rho_u + \rho_d + \rho_s)/3$ and with the nuclear saturation density $\rho_0 = 0.17 \text{fm}^{-3}$. This is shown in fig. \ref{grafEpart}, where we can identify two stability lines for pure phases (each beginning at each first-order transition) and two mixed phase regions whose description has not been considered. The minimum, which corresponds to the beginning of the first stable pure phase, occurs for lower values of $E/A$ and $\rho / \rho_0$ than in simpler versions of the NJL model (for example, in~\cite{Buballa:2003qv}, the minumum is located at $\rho = 2.25 \rho_0$ and $E/A = 1102 \text{MeV}$). This is also true for the onset point of a finite strange quark density (in the same example~\cite{Buballa:2003qv}, this occurs at $\rho = 3.85 \rho_0$ and $E/A \sim 1140 \text{MeV}$). However, the minimum is still at a higher value of $E/A$ than that of stable nuclear matter ($\sim 930 \text{MeV}$).

Finally, we inspect the $p \left(\epsilon\right)$ equation of state of the model, which is given in fig. \ref{grafpEner}. This is required for the integration of the Tolman-Oppenheimer-Volkoff equations
%~\cite{tov1,tov2}
 which are used in the description of compact stars. The equation of state we have obtained is relatively soft when compared with a number of other proposals~\cite{Weber:2004kj}, which results in a maximum stellar radius of $\sim 10.4 \text{km}$ and a maximum mass of $\sim 1.54 \text{M}_{\odot}$. The latter value does not conform to recent observations of stars with masses $\sim 2 \text{M}_{\odot}$~\cite{Demorest:2010bx,Antoniadis:2013pzd}, which should not come as a surprise the model does not contemplate yet relevant features to compact stars, such as the magnetic field (which should be high in compact stars, with a significant impact on the equation of state) as well as spin 1 interactions and rotational effects.

\section{Conclusion}

The main result which should be emphasized is the appearance of a second first-order boundary associated with a transition in the dynamical mass of the strange quarks. This striking new feature is related with the ordering $m_{\kappa} < m_{a_0}$ in the values of the low lying scalar meson masses. The $g_3$ parameter, which corresponds to a non-OZI violating term and is very sensitive to the above mentioned mass values, appears to counterbalance the effects of the flavour-mixing 't Hooft term which tends to smooth out the second transition.

Regarding the implications for SQM, our results show values at the minimum ($\rho \sim 1.6 \rho_0$ and $E/A \sim 975 \text{MeV}$) and at the onset of SQM ($\rho \sim 3.3 \rho_0$ and $E/A \sim 1042 \text{MeV}$) which are lower than those reported in the literature for previous versions of the NJL model. These values are above nuclear matter stability, thus excluding the idea of absolutely stable SQM, but are still more favourable for its formation at higher densities.

A more detailed discussion which includes a description of the mixed phases can be found in~\cite{Moreira:2014qna}.

Research supported by Centro de F\'{i}sica Computacional of the University of Coimbra,  by the Funda\c{c}\~ao para a Ci\^{e}ncia e Tecnologia grant SFRH/BPD/63070/2009, Or\c{c}amento de Estado and by the European Community - Research Infrastructure Integrating Activity Study of Strongly Interacting Matter (Grant Agreement 283286) under the Seventh Framework Programme of EU.

% % % % % % % % % % % % % % % % % % % % % % % % % % % % % % % % % % % % % % % % % % % % % % % % % % %

\bibliographystyle{h-elsevier3}

\bibliography{Bibliography}

\end{document}